\documentclass[acmtoms]{acmtrans2m}

\acmVolume{V}
\acmNumber{N}
\acmYear{YY}
\acmMonth{Month}

\usepackage{amsmath}
\usepackage{color}
\usepackage{graphicx}
\usepackage{booktabs}
\usepackage{colortbl}
\usepackage{enumerate}

\usepackage[dotinlabels]{titletoc}
\usepackage{array}
\usepackage{soul}
\usepackage{moreverb}
\usepackage{rotating}

\newcommand{\algname}{\textcolor{white}{ (NAME)}}

\usepackage{listings}
\usepackage{caption}
\DeclareCaptionFont{white}{\color{white}}
\DeclareCaptionFont{black}{\color{black}}
\DeclareCaptionFormat{listing}{\colorbox[cmyk]{0.23, 0.15, 0.15,0.01}{\parbox{\textwidth}{\centering #1\algname{}#2#3}}}
\title{Solving Sequences of Generalized Least-Squares Problems on Multi-threaded Architectures}
            
\author{
DIEGO FABREGAT-TRAVER \\
Aachen Institute for Advanced Study in Computational Engineering Science, \\RWTH Aachen
\and
YURII AULCHENKO \\
Institute of Cytology and Genetics, SD RAS
\and
PAOLO BIENTINESI \\
Aachen Institute for Advanced Study in Computational Engineering Science, \\RWTH Aachen
}
            
\begin{abstract} 
Generalized linear mixed-effects models in the context of
genome-wide association studies (GWAS) represent a formidable computational challenge:
the solution of millions of correlated generalized least-squares problems, and the processing of terabytes of data.
We present high performance in-core and out-of-core shared-memory algorithms for GWAS:
By taking advantage of domain-specific knowledge, 
exploiting multi-core parallelism, and handling data efficiently, our algorithms attain
unequalled performance. When compared to GenABEL, one of the most widely used libraries for GWAS, on a 12-core processor
we obtain 50-fold speedups. As a consequence, our routines enable genome studies of unprecedented size.
\end{abstract}



            
\begin{document}
            
\begin{bottomstuff} 
Authors' addresses:
Diego Fabregat, AICES, RWTH Aachen, Aachen, Germany, {\tt fabregat@aices.rwth-aachen.de}. % \newline
Yurii Aulchenko, Institute of Cytology and Genetics, SD RAS, Novosibirsk, Russia, {\tt yurii.aulchenko@gmail.com}. % \newline
Paolo Bientinesi, AICES, RWTH Aachen, Aachen, Germany, {\tt pauldj@aices.rwth-aachen.de}. % \newline
\end{bottomstuff}
            
\maketitle

\section{Introduction}
\label{sec:intro}

Generalized linear mixed-effects models (GLMMs) are a type of statistical model widespread in many different disciplines such
as genomics, econometrics, and social sciences~\cite{Teslovich2010-short,antonio200758,ISI:000277906700004}. 
Applications based on GLMMs face two computational challenges:
the solution of a sequence comprising millions of generalized least-squares problems (GLSs), and
the processing of data sets so large that they only fit in secondary storage devices.
In this paper, we target the computation of GLMMs in the context 
of genome-wide association studies (GWAS).

GWAS is the tool of choice to analyze the relationship
between DNA sequence variations and complex traits such as 
diabetes and coronary heart diseases~\cite{10.1371/journal.pgen.1001256-short,Levy2009-short,Speliotes2010-short}. 
More than 1400 papers published during the last five years endorse the relevance of GWAS~\cite{GWAScatalog}.
The GLMM specific to GWAS solves the equation
\begin{equation}
\label{eq:probDef}
%\left\{ 
{\begin{aligned}
    b_{i} & := (X_i^T M^{-1} X_i)^{-1} X_i^T M^{-1} y,
    %M     & := h \Phi + (1 - h) I
\end{aligned}}
%\right.
\;
{\begin{aligned}
		\text{ \hspace{-1.8mm} with } & 1 \le i \le m%, \text{ and } \\
%              & 1 \le j \le t
\end{aligned}}
\end{equation}
where $y$ is the vector of observations, representing a given trait or phenotype; $X_i$ is the design matrix, 
including covariates and genome measurements; $M$ represents dependencies among observations; and $b_{i}$
represents the relation between a variation in the genome sequence ($X_i$) and a variation in the trait ($y$).
In linear algebra terms, Eq.~\eqref{eq:probDef} solves a linear regression with non-independent outcomes where
$b_{i} \in \mathcal{R}^{p}$, 
$X_i \in \mathcal{R}^{n \times p}$ is full rank, 
$M \in \mathcal{R}^{n \times n}$ is symmetric positive definite (SPD), and
$y \in \mathcal{R}^{n}$;
%$\Phi \in \mathcal{R}^{n \times n}$ is symmetric,
%$I \in \mathcal{R}^{n \times n}$, and 
%$h \in \mathcal{R}$; 
the sizes are as follows:
$n \approx 10^4$, 
$1 \le p \le 20$,  and 
$m$, the length of the sequence, ranges from $10^6$ to $10^8$.
The quantities $X_i$, $M$, and $y$ are known.
Additionally, the $X_i$'s present a special structure that will prove to be critical for performance:
each $X_i$ may be partitioned as $(X_L \; | \; X_{R_i})$, where $X_L$ is the same for all $X_i$'s.

\subsection{Limitations}
Computational biologists performing GWAS aim for the sizes described above;
in a typical scenario, 3 Terabytes of data have to be processed through $3.6 \times 10^{15}$
arithmetic operations (Petaflops).
In practice, current GWAS solvers are constrained to much smaller problems due to
time limitations. For instance, in~\cite{20233392}, the authors carry out
a study that takes almost 4 hours for the following problem sizes: $n = 1{,}500$, 
$p = 4$, and $m = 220{,}833$. The time to perform the same study for $m = 2.5$ million 
is estimated to be roughly 43 hours. With our routines, the time to complete the latter
reduces to 10 minutes. 

\subsection{Terminology}

We collect and give a brief description of the acronyms used throughout the paper.

\begin{itemize}
	\item {\sc gwas}: Genome-Wide Association Studies
	\item {\sc gls}: Generalized Least-Squares problems
	\item GenABEL: One of the most widely used frameworks to perform {\sc gwas}
	\item {\sc gwfgls}: GenABEL's state-of-the-art routine for the solution of Eq.~\eqref{eq:probDef}
	\item {\sc hp-gwas}: our novel in-core solver for Eq.~\eqref{eq:probDef}
	\item {\sc ooc-hp-gwas}: out-of-core version of {\sc hp-gwas}.
\end{itemize}

Table~\ref{tab:blas-lapack} enumerates the BLAS (Basic Linear Algebra Subprograms)~\cite{BLAS3}
and LAPACK (Linear Algebra PACKage)~\cite{laug} routines used in the algorithms presented in this paper.
LAPACK and BLAS are the de-facto standard libraries for high-performance dense linear algebra computations. 

\begin{table*}
\renewcommand{\arraystretch}{1.4}
\caption{}
\label{tab:blas-lapack}
\centering
  \begin{tabular}{l@{\hspace*{8mm}} l@{\hspace*{8mm}} c} \toprule
	  \multicolumn{3}{c}{\bf BLAS 1 and 2} \\ \midrule
	  {\sc dot} & Dot product & $\alpha := x^T y$ \\
	  {\sc gemv} & Matrix-vector product & $y := A x + y$ \\
	  {\sc trsv} & Triangular system with single right-hand side & $A x = b$ \\
	  \bottomrule
	  \multicolumn{3}{c}{\bf BLAS 3} \\ \midrule
	  {\sc gemm} & Matrix-matrix product & $C := A B + C$ \\
	  {\sc syrk} & Rank-k update & $C := A^T A + C$ \\
	  {\sc trsm} & Triangular system with multiple right-hand sides & $A X = B$ \\
	  \bottomrule
	  \multicolumn{3}{c}{\bf LAPACK} \\ \midrule
	  {\sc getri} &\multicolumn{2}{l}{Inversion of a general matrix} \\
	  {\sc gesv}  &\multicolumn{2}{l}{General system with multiple right-hand sides } \\
	  {\sc posv}  &\multicolumn{2}{l}{SPD system with multiple right-hand sides} \\
	  {\sc potrf} &\multicolumn{2}{l}{Cholesky factorization} \\
	  {\sc gels}  &\multicolumn{2}{l}{Solution of a least-squares problem} \\
	  {\sc ggglm} &\multicolumn{2}{l}{Solution of a general Gauss-Markov linear model} \\
		\bottomrule

  \end{tabular}
\end{table*}

\subsection{Related Work}

Traditionally, LAPACK is the tool of choice to develop high-performance 
algorithms and routines for linear algebra operations.
Although LAPACK does not support the solution of a single
GLS directly, it offers routines for closely
related problems:
{\sc gels} for least squares problems, and
{\sc ggglm} for the general Gauss-Markov linear model. 
Algorithms~\ref{alg:gels}~and~\ref{alg:ggglm} provide examples for the reduction of GLS problems 
to {\sc gels} and {\sc ggglm}, respectively.
Unfortunately, none of the algorithms provided by LAPACK is able to exploit the sequence 
of GLSs within GWAS, nor the specific structure of its operands. 
Conversely, existing ad-hoc routines for Eq.~\eqref{eq:probDef}, such as the widely used {\sc gwfgls}, 
are aware of the specific knowledge arising
from the application, but exploit it in a sub-optimal way.

LAPACK and GenABEL present additional drawbacks:
LAPACK routines are in-core, i.e., data must fit in main memory;
since GWAS may involve terabytes of data, it is in general 
not feasible to use these routines directly.
Contrarily, GenABEL incorporates an 
out-of-core mechanism, but it suffers from significant overhead.

\begin{center}
\renewcommand{\algname}{}
\renewcommand{\lstlistingname}{Algorithm}
\begin{minipage}{0.85\linewidth}
\begin{lstlisting}[caption=GLS problem reduced to {\sc gels}, escapechar=!,label=alg:gels]
 $L L^T = M$                (!\sc potrf!)
 $y := L^{-1} y$                (!\sc trsv!)
 $X := L^{-1} X$                (!\sc trsm!)
 $b := $!\sc gels!$(X, y)$
\end{lstlisting}
\end{minipage}
%\hspace*{10mm}
\begin{minipage}{0.85\linewidth}
\begin{lstlisting}[caption=GLS problem reduced to {\sc ggglm}, escapechar=!,label=alg:ggglm]
 $L L^T = M$                (!\sc potrf!)
 $b := $!\sc ggglm!$(X, y, L)$
\end{lstlisting}
%$b := ggglm(X, y, L)$ !\vspace*{8.4mm}!
\end{minipage}
\end{center}

\subsection{Contributions}

We present high-performance in-core and out-of-core algorithms, {\sc hp-gwas} and {\sc ooc-hp-gwas},
and their corresponding routines for the computation of GWAS on multi-threaded architectures. 

Our algorithms are optimized not for a single instance of the GLS problem 
but for the whole sequence of such problems. This is accomplished by
\begin{itemize}
\item breaking the black box structure of traditional libraries, which impose a 
	separate routine call for each individual GLS,
\item exploiting domain-specific knowledge such as the particular structure of the operands,
\item grouping successive problems, allowing the use of high performance kernels at their
	full potential, and
\item organizing the computation to use multiple types of parallelism.
\end{itemize}
When combined, these optimizations lead to an in-core routine that outperforms
GenABEL's {\sc gwfgls} by a factor of 50.

Additionally, we enable the solution of very large sequences of problems by incorporating
an efficient out-of-core mechanism to our in-core routine. 
Thanks to this extension, the out-of-core
routine is capable of sustaining the high performance of the in-core one for data sets
as large as the secondary storage.

\subsection{Organization}

Section~\ref{sec:incore} details, through a series of improvements,
how {\sc hp-gwas} exploits both domain-specific knowledge and the BLAS library 
to attain high performance and scalability.
In Section~\ref{sec:results-incore}, we quantify the gain of each improvement
and present a performance comparison between {\sc hp-gwas} and {\sc gwfgls}.
Section~\ref{sec:ooc} exposes the key ideas behind the out-of-core mechanism 
leading to {\sc ooc-hp-gwas}, which maintains {\sc hp-gwas} performance 
for very large sets of data. Out-of-core results are provided
in Section~\ref{sec:results-ooc}. We discuss future work in Section~\ref{sec:future},
and draw conclusions in Section~\ref{sec:conclusions}.

\section{In-core algorithm}
\label{sec:incore}

We commence the discussion by describing the incremental steps to transform a generic algorithm
for the solution of a single GLS problem into a high-performance algorithm that
1) solves a sequence of GLS problems, 
2) exploits GWAS-specific knowledge, and 
3) exploits multi-core parallelism.
The resulting algorithm is then used in Section~\ref{sec:ooc} as a starting point
towards a high-performance out-of-core algorithm.

Algorithm~\ref{alg:gls-one} solves a generic GLS problem. The approach consists
in first reducing the GLS to a linear least-squares problem (as shown in
Algorithm~\ref{alg:gels}), and then solving
the associated normal equations $(X^T X)^{-1} X^T y$, where the coefficient matrix
$X \in R^{n \times p}$ is full rank and $n > p$.
To this end, Algorithm~\ref{alg:gls-one} first factors $M$ via a Cholesky
factorization: $L L^T = M$; and then, it solves the systems $X := L^{-1} X$ and $y := L^{-1} y$.
Several alternatives exist for the solution of the normal equations; for
a detailed discussion we refer the reader to~\cite{Golub:1996:MC:248979,bjor:96}.
Numerical considerations allow us to safely rely on the Cholesky factorization of the SPD matrix
$S := X^T X$ without incurring instabilities. The algorithm completes by computing $b := X^T y$ and solving the linear system $b = S^{-1} b$.
For each operation in the algorithm,
we specify in brackets the corresponding BLAS/LAPACK routine.

\begin{center}
\renewcommand{\algname}{\textcolor{black}{$\ $({\sc black-box})}}
\renewcommand{\lstlistingname}{Algorithm}
\begin{minipage}{0.90\linewidth}
\begin{lstlisting}[caption=Solution of a GLS problem, escapechar=!,label=alg:gls-one]
 $L L^T = M$                (!\sc potrf!)
 $X := L^{-1} X$                (!\sc trsm!)
 $y := L^{-1} y$                (!\sc trsv!)
 $S := X^T X$                (!\sc syrk!)
 $b := X^T y$                (!\sc gemv!)
 $b := S^{-1} b$                (!\sc posv!)
\end{lstlisting}
\end{minipage}
\end{center}

Algorithm~\ref{alg:gls-one} solves a single GLS problem. The algorithm may be used 
to solve a sequence of problems in a black box 
fashion, i.e., for each individual coefficient matrix $X_i$, use Algorithm~\ref{alg:gls-one} to
solve the corresponding GLS problem. As the reader might have noticed, this approach leads
to a considerable amount of redundant computation. We avoid the black box approach, and exploit
the fact that we are solving a sequence of correlated problems. A closer look at Algorithm~\ref{alg:gls-one}
reveals that, since only $X$ varies from problem to problem, operations at lines 1 and 3 may
be performed once and reused across the sequence. The resulting Algorithm~\ref{alg:gls-seq-one}
greatly reduces the computation performed using a black box approach.

\begin{center}
\renewcommand{\algname}{\textcolor{black}{$\ $({\sc seq-gls})}}
\renewcommand{\lstlistingname}{Algorithm}
\begin{minipage}{0.90\linewidth}
\begin{lstlisting}[caption=Solution of a sequence of GLSs, escapechar=!,label=alg:gls-seq-one]
 $L L^T = M$                 (!\sc potrf!)
 $y := L^{-1} y$                 (!\sc trsv!)
 for each $X_i$
   $X_i := L^{-1} X_i$                 (!\sc trsm!)
   $S_i := X_i^T X_i$                 (!\sc syrk!)
   $b_i := X_i^T y$                 (!\sc gemv!)
   $b_i := S_i^{-1} b_i$                 (!\sc posv!)
\end{lstlisting}
\end{minipage}
\end{center}

Although Algorithm~\ref{alg:gls-seq-one} already solves
a sequence of GLS problems, it is still sub-optimal in a number of ways. 
The first crucial step towards high performance is a reorganization of
the computation.
A large percent of the computation in the loop is carried out by the {\sc trsm} operation
at line 4. Even though {\sc trsm} is a BLAS-3 operation, the fact that the
system is solved for, at most, 20 right-hand sides does not allow {\sc trsm}
to reach its peak performance; thus, the overall performance is
affected. To overcome this limitation, we take advantage again 
from the sequence of problems: we group multiple {\sc trsm}s corresponding
to successive problems $L^{-1} X_i$ into a larger {\sc trsm} with enough right-hand
sides to deliver its maximum performance, i.e., $L ^{-1}\mathcal{X}$, where $\mathcal{X}$
represents the collection of all $X$'s: ($ X_1 \,| \, X_2 | \, \ldots \, | \, X_m$).

As a further improvement, we focus on the knowledge specific to GWAS:
the special structure of $X$. 
Each individual $X_i$ may be partitioned in $(X_L \, | \, X_{R_i})$, where $X_L$ is fixed;
thus the {\sc trsm} operation $L ^{-1}\mathcal{X}$ may be split into two {\sc trsm}s:
$L^{-1} X_L$ and $L^{-1} \mathcal{X}_R$, where $\mathcal{X}_R$
represents the collection of all $X_R$'s: ($ X_{R_1} \, | \, X_{R_2} \, | \, \ldots \,| \, X_{R_m}$).
Additionally, the fact that $X_L$ is fixed allows for more computation reuse: 
as shown in Fig.~\ref{fig:part}, the top left part of $S_i$, and the top part of $b_i$ are also fixed.
The resulting algorithm is assembled in Algorithm~\ref{alg:inc-gwas}, {\sc hp-gwas}.

\begin{figure}
  \centering
	$
        \renewcommand{\arraystretch}{1.4}
		\left( {\begin{array}{@{}c|c@{}}
			S_{TL} = X_L^T X_L & S_{TR} = X_L^T X_R \\\hline
			S_{BL} = X_R^T X_L & S_{BR} = X_R^T X_R \\
		\end{array}} \right)
		\hspace{5mm}
		;
		\hspace{5mm}
		\left( {\begin{array}{@{}c@{}}
			b_{T} = X_L^T y_T \\\hline
			b_{B} = X_R^T y_B \\
		\end{array}} \right)
    $
\caption{Computation of $S := X^T X $, and $b := X^T y$  in terms of the parts of $X$: $(X_L \, | \, X_R)$.
$L$, $R$, $T$, and $B$, stand for $L$eft, $R$ight, $T$op, and $B$ottom, respectively.} 
\label{fig:part}
\end{figure}

\begin{center}
\renewcommand{\algname}{\textcolor{black}{$\ $({\sc hp-gwas})}}
\renewcommand{\lstlistingname}{Algorithm}
%\begin{minipage}{0.40\linewidth}
%\begin{lstlisting}[caption=Solution of a sequence of GLSs - Blocked, escapechar=!,label=alg:gls-seq-two]
%$L L^T = M$                 (!\sc potrf!)
%$X := L^{-1} X$                 (!\sc trsm!)
%$y := L^{-1} y$                 (!\sc trsv!)
%for each $X_i$
  %$S := X_i^T X_i$                 (!\sc syrk!)
  %$b_i := X_i^T y$                 (!\sc gemv!)
  %$b_i := S^{-1} b_i$                 (!\sc posv!) !\vspace*{16.44mm}!
%\end{lstlisting}
%\end{minipage}
%\hfill
\begin{minipage}{0.90\linewidth}
\begin{lstlisting}[caption=Solution of the GWAS-specific sequence of GLSs, escapechar=!,label=alg:inc-gwas]
 $L L^T = M$                   (!\sc potrf!)
 $X_L := L^{-1} X_L$                   (!\sc trsm!)
 $\mathcal{X}_R := L^{-1} \mathcal{X}_R$                   (!\sc trsm!)
 $y := L^{-1} y$                   (!\sc trsv!)
 $S_{TL} := X_L^T X_L$                   (!\sc syrk!)
 $b_T := X_L^T y$                   (!\sc gemv!)
 for each $X_{R_i}$
   $S_{BL_i} := X_{R_i}^T X_L$                    (!\sc gemv!)
   $S_{BR_i} := X_{R_i}^T X_{R_i}$                    (!\sc dot!)
   $b_{B_i} := X_{R_i}^T y$                    (!\sc dot!)
   $b_i := S_i^{-1} b_i$                    (!\sc posv!)
\end{lstlisting}
\end{minipage}
\end{center}
   %$b_i := 
        %\renewcommand{\arraystretch}{1.2}
		%\left( {\begin{array}{@{}c@{\,}|@{\,}c@{}} 
				%S_{TL} & S_{BL_i}^T  \\\hline
				%S_{BL} & S_{BR_i}
        %\end{array}} \right)
        %\left( {\begin{array}{@{}c@{}} 
				%b_{T}^T  \\\hline
				%b_{B_i}
        %\end{array}} \right)
	%$

\subsection{GenABEL's {\sc gwfgls}}

For completeness we provide in Algorithm~\ref{alg:gwfgls} the algorithm implemented by 
GenABEL's {\sc gwfgls} routine. The algorithm takes advantage from the specific 
structure of GWAS by computing lines 1 and 2 once, and reusing the results 
across the sequence of problems.
Unfortunately, a number of choices prevent it from attaining high performance:
\begin{itemize}
\item the inversion of $M$ (line 1) performs 6 times more computation than a 
	Cholesky factorization of $M$,
\item line 2 breaks the symmetry of the expression $(X^T M^{-1} X)^{-1}$,
	which translates into doubling the amount of computation performed,
\item the BLAS-2 operation at line 4 ({\sc gemv}) could be cast as a single
	BLAS-3 {\sc gemm} involving all $X_R$'s (what we called $\mathcal{X}_R$
	in Algorithm~\ref{alg:inc-gwas}); {\sc gwfgls} does not include this improvement, 
	thus it does not benefit from {\sc gemm}'s high performance.
\end{itemize}

\begin{center}
\renewcommand{\algname}{\textcolor{black}{$\ $({\sc gwfgls})}}
\renewcommand{\lstlistingname}{Algorithm}
\begin{minipage}{0.90\linewidth}
\begin{lstlisting}[caption=GenABEL's algorithm for GWAS, escapechar=!,label=alg:gwfgls]
 $M = M^{-1}$                   (!\sc getri!)
 $W_L^T := X_L^T M$                   (!\sc gemm!)
 for each $X_{R_i}$
   $W_{R_i}^T := X_{R_i}^T M$                   (!\sc gemv!)
   $S_i := W^T X_i$                   (!\sc gemm!)
   $b_i := W^T y$                   (!\sc gemv!)
   $b_i := S^{-1} b_i$                   (!\sc gesv!)
\end{lstlisting}
\end{minipage}
\end{center}
%!\vspace*{18.44mm}!

\subsection{Computational cost}
\label{subsec:cost}

Table~\ref{tab:cost} includes the asymptotic cost of Algorithms~\ref{alg:gls-one}~--~\ref{alg:gwfgls} together with
the ratio over our best algorithm, {\sc hp-gwas}.
A discussion of the provided data follows.

\begin{table*}
\renewcommand{\arraystretch}{1.4}
\centering
  \begin{tabular}{l@{\hspace*{8mm}} c@{\hspace*{8mm}} c} \toprule
	  & {Computational cost} & Ratio {\sc Alg.\#} / {\sc Alg.~\ref{alg:inc-gwas}} \\ \midrule
	  {\bf {\phantom{y}{\sc Alg.~\ref{alg:gls-one} (black-box)}\phantom{y}} }   & $\approx n$ & $O(m n^3)$ \\[2mm]
		{\bf {\phantom{y}{\sc Alg.~\ref{alg:gls-seq-one} (seq-gls)}\phantom{y}} } & $\approx p$ & $O(n^3 + m n^2 p + m n p^2)$ \\[2mm]
	    {\bf {\phantom{y}{\sc Alg.~\ref{alg:inc-gwas} (hp-gwas)}\phantom{y}} }    & $1$         & $O(n^3 + m n^2 + m n p)$     \\[2mm]
		{\bf {\phantom{y}{\sc Alg.~\ref{alg:gwfgls} (gwfgls)} \phantom{y} } }     & $\approx 2$ & $O(n^3 + m n^2 + m n p^2)$   \\[2mm]
	\bottomrule
  \end{tabular}
\caption{Asymptotic cost of each of the presented algorithms for GWAS. %(Algs.~\ref{alg:gls-one}~--~\ref{alg:gwfgls}).
The ratio over {\sc hp-gwas} shows the progressive improvement made from the initial black box approach.
{\sc hp-gwas} also improves the cost of {\sc gwfgls} by a constant factor.}
\label{tab:cost}
\end{table*}

\begin{enumerate}
	\item The solution of a single GLS problem via Algorithm~\ref{alg:gls-one}, {\sc black-box}, has a computational
cost of $O(n^3)$. The solution of a sequence of such problems using this algorithm as a
black box entails thus $O(m n^3)$ flops, corresponding to the 
computation of $m$ Cholesky factorizations. 
Clearly, this is not the best approach to solve a sequence of correlated problems:
it performs $n$ times more operations than {\sc hp-gwas}.
\item The key insight in Algorithm~\ref{alg:gls-seq-one}, {\sc seq-gls}, is to
take advantage from the fact that we are solving not one but a sequence of 
correlated problems.
Based on an analysis of dependencies, the algorithm breaks the rigidity of {\sc black-box},
and rearranges the computation. As a result, the computational cost is reduced by a factor of $n/p$. 
Even though {\sc seq-gls} represent a great improvement with respect to {\sc black-box},
it is still not optimal for GWAS: a further reduction of redundant computation is possible.
\item {\sc hp-gwas} incorporates two further optimizations to overcome
the limitations of {\sc seq-gls}. First,
the algorithm exposes the structure of $X$, and the quantities computed
from it, completely eliminating redundant operations. Then, the computation
is carefully reorganized to exploit the full potential of the underlying
libraries, resulting in an extremely efficient algorithm (see Fig.~\ref{fig:incore-one}).
\item As for {\sc gwfgls}, it benefits from both the sequence of problems and the
specific structure of $X$. Unfortunately, the algorithm fails at exploiting the existing
symmetries, thus performing twice as much computation as {\sc hp-gwas}.
Additionally, the algorithm is not properly designed to benefit from the highly-optimized
BLAS library, having a negative impact on its performance.
\end{enumerate}

\subsection{Parallelism}

{\sc hp-gwas} relies on a set of kernels provided by the highly-optimized BLAS
and LAPACK libraries. In this situation, a straightforward approach to target multi-core
architectures is to link the routine to a multi-threaded version of the libraries.
While the first section of {\sc hp-gwas} (lines 1 to 6) benefits from this approach,
showing high scalability, the second section (lines 7 to 11) does not scale.
Therefore the weight of the second, although small in the sequential case, increases with
the number of cores, affecting the overall scalability.
To address this shortcoming we use a different parallelization scheme for the two sections:
multi-threaded BLAS for lines 1 to 6, and OpenMP parallelism with single-threaded BLAS
for lines 7 to 11. As we show in the next section, the resulting routine is highly scalable.

\section{Performance results (I)}
\label{sec:results-incore}

We turn now the attention towards the experimental results. We first report
on timings for all four presented algorithms %(Algs.~\ref{alg:gls-one}~--~\ref{alg:gwfgls})
for the sequential case; the goal is to show and discuss the effect of the improvements described
in the previous section. Then we focus on {\sc hp-gwas} and {\sc gwfgls}; we concentrate
on timings for the multi-threaded versions of the routines and their scalability.

\subsection{Experimental setup}
As a computing environment we chose an architecture that we believe is readily
available to most computational scientists.
All four algorithms were implemented in C. Although GenABEL's interface is written in R,
{\sc gwfgls} and most of its routines are written in C.
We ran all tests on a SMP system made of two Intel Xeon X5675 multi-core processors. 
Each processor has six cores, operating at a frequency of 3.06 GHz, for
a combined peak performance of 146.88 GFlops/sec.
The system is equipped with 32GB of RAM and 1TB of disk as secondary memory.
We compiled the routines with the GNU C Compiler (gcc, version 4.4.5), and
linked to a multi-threaded Intel's MKL library (version 10.3).
{\sc hp-gwas} also makes use of the OpenMP parallelism provided by the compiler through a number of {\it pragma} directives.

\begin{figure*}
\centering
%\begin{minipage}[t]{0.6\textwidth}
%\vspace{0pt}
\includegraphics[scale=0.65]{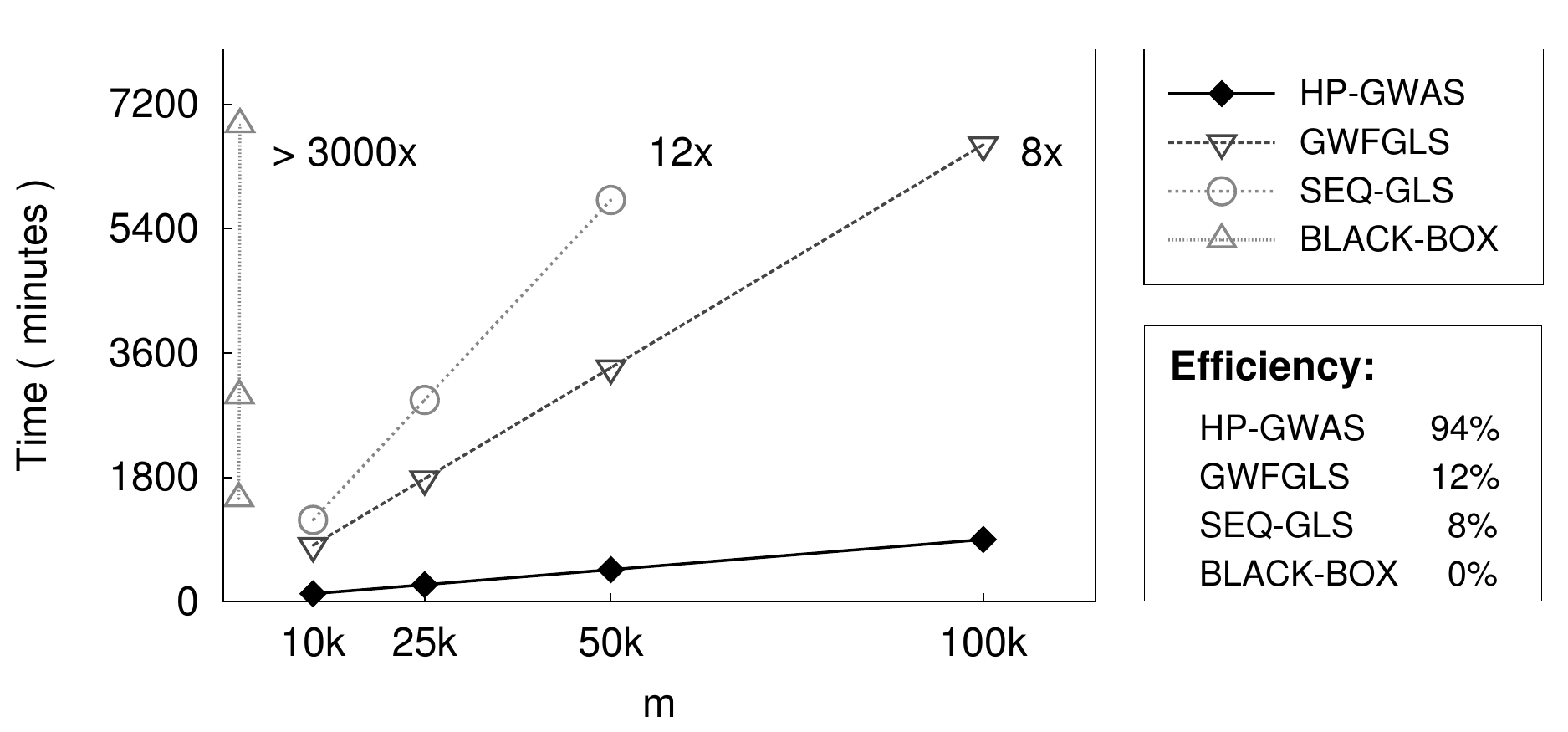}
%\makeatletter\def\@captype{figure}\makeatother
\caption{Comparison of the four presented algorithms for GWAS. The gain of each
	individual improvement from {\sc black-box} to {\sc hp-gwas} is illustrated.
	Additionally, our best algorithm, {\sc hp-gwas}, outperforms state-of-the-art
	{\sc gwfgls} by a factor of 8. All experiments were performed using a single thread.
	The other problem dimensions are: $n=10{,}000$, and $p=4$.} 
\label{fig:incore-one}
%\end{minipage}
\end{figure*}

\subsection{Results}

Fig.~\ref{fig:incore-one} shows the timings of all four algorithms
for an increasing value of $m$, the number of GLS problems to be solved.
The experiments were run using a single thread.
The results for {\sc black-box} exemplify the limitations of 
solving a sequence of correlated problems as if they are unrelated: %independent problems:
no matter how optimized the algorithm is for a single instance, it cannot
compete with algorithms specially tailored to solve the sequence as a whole.
As a first step towards high performance, {\sc seq-gls} reuses computation across
the sequence of problems. Consequently, the algorithm reduces dramatically 
the execution time of the naive {\sc black-box} approach, leading to a speedup 
greater than 250.

{\sc hp-gwas} further reduces the execution time of {\sc seq-gls} by a factor of 12.
The gain is explained by the effect of two optimizations. 
On the one hand, {\sc hp-gwas} exploits 
application-specific knowledge, the structure of $X$, leading to a
speedup of $p = 4$ (larger values of $p$ result in even larger speedups). On the other
hand, the computation is reorganized taking into account high-performance considerations.
It is a common misconception that every BLAS routine attains the same efficiency.
However, due to architectural constraints such as memory hierarchy and associated latency,
BLAS-3 routines attain higher efficiency than BLAS-1 and BLAS-2. Therefore,
rewriting multiple {\sc trsv}s (BLAS-2) as a single large {\sc trsm} (BLAS-3),
our algorithm achieves an extra speedup of 3.
As shown in Fig.~\ref{fig:incore-one}, {\sc hp-gwas} is an efficient
algorithm to carry out GWAS; it  attains 94\% of the architecture's peak performance.

Although {\sc gwfgls} is aware of the specific properties
of GWAS and benefits from such knowledge, the algorithm suffers from 
inefficiencies similar to {\sc seq-gls}: it still performs redundant computation,
and it is not properly tailored to benefit from BLAS-3 performance.
The combination of both shortcomings results in a routine that is
8 times slower than {\sc hp-gwas}.

\begin{figure*}
\centering
%\begin{minipage}[t]{0.60\textwidth}
%\vspace{0pt}
\includegraphics[scale=0.85]{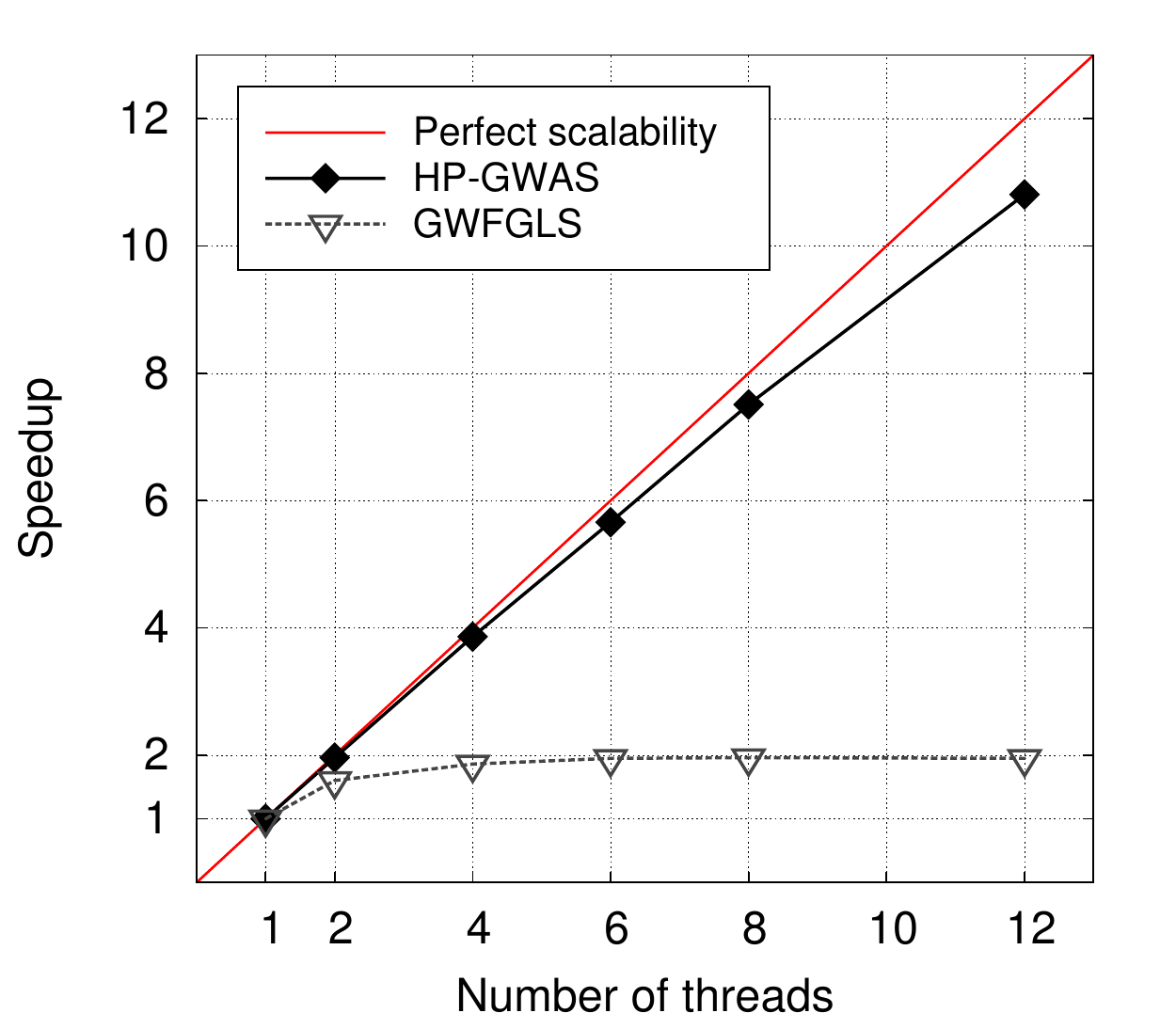}
%\makeatletter\def\@captype{figure}\makeatother
\caption{Scalability of {\sc hp-gwas} and {\sc gwfgls}. While {\sc gwfgls}'
speedup plateaus at 2, and the gain is minimal for more than 4 cores, 
{\sc hp-gwas} attains high-scalability and an even larger
speedup is foreseen for a greater number of cores.
The problem dimensions are: $n=10{,}000$, $p=4$, and $m=100{,}000$.} 
\label{fig:scalability}
%\end{minipage}
\end{figure*}

Henceforth, we concentrate on {\sc hp-gwas} and {\sc gwfgls}.
In Fig.~\ref{fig:scalability} we report on the scalability of both algorithms.
As the figure reflects, while {\sc gwfgls} barely reaches a speedup of 2, completely
stalling after 6 cores are used, {\sc hp-gwas} attains a speedup of almost 11 when 
using 12 cores. Most interestingly, the tendency clearly shows that larger speedups
are expected for {\sc hp-gwas} when increasing the number of cores available.

The disparity in the scalability of these two algorithms is mainly due to their use
of the BLAS library. In the case of {\sc gwfgls}, the algorithm casts most of the
computation in terms of the BLAS-2 operation {\sc gemv}, which, being a memory-bound
operation, is limited not only in performance but also in scalability. Instead,
as described earlier, {\sc hp-gwas} mainly builds on top of {\sc trsm} (BLAS-3),
which attains high scalability when operating on a large number of right-hand sides.

\begin{figure*}
\centering
%\begin{minipage}[t]{0.60\textwidth}
%\vspace{0pt}
\includegraphics[scale=0.82]{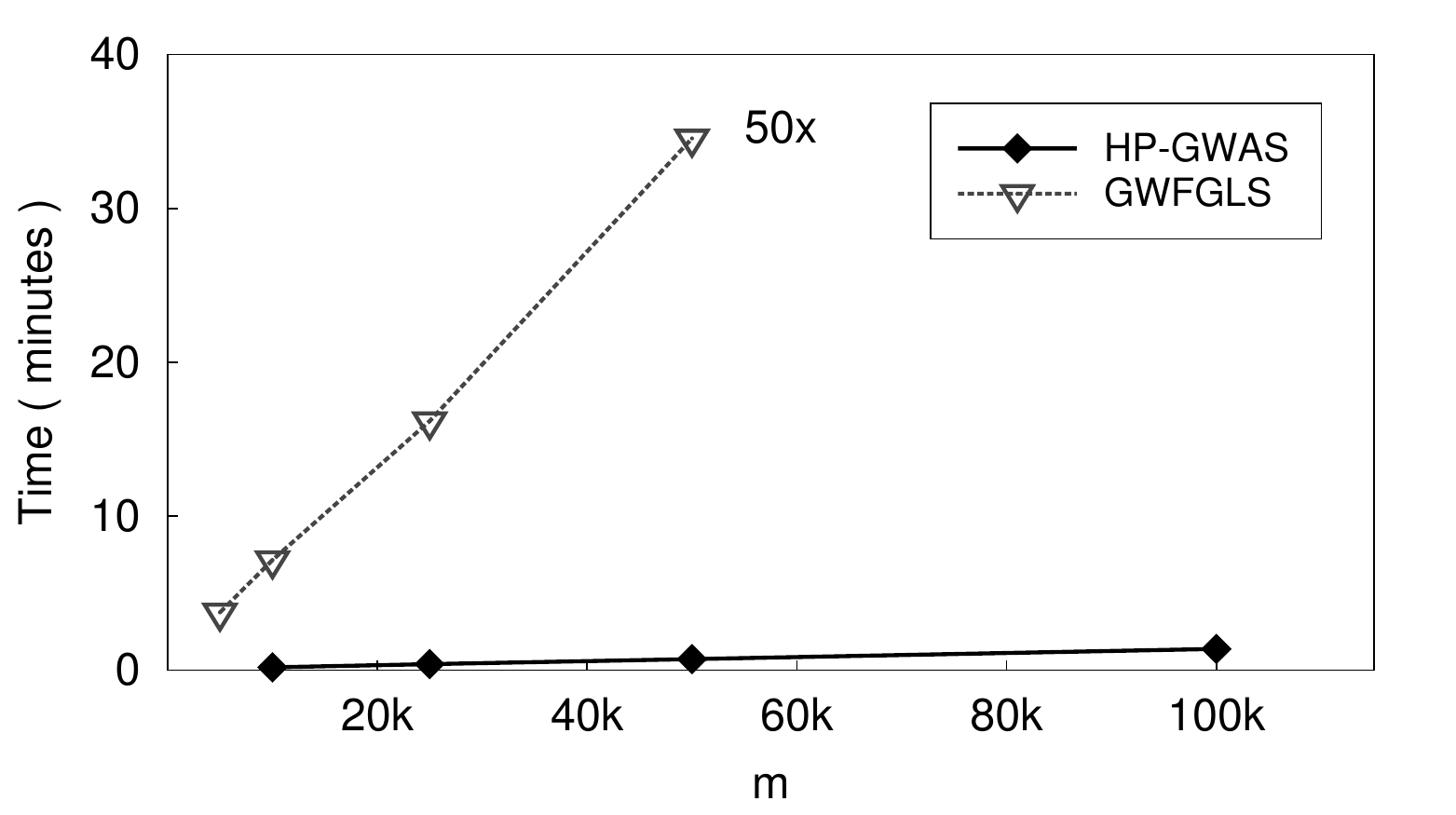}
%\makeatletter\def\@captype{figure}\makeatother
\caption{Timings for the multi-threaded versions of {\sc hp-gwas} and {\sc gwfgls}.
Thanks to a much better scalability, our routine outperforms {\sc gwfgls} by a factor of 50.
The experiments were run using 12 threads. The other problem dimensions are: $n=10{,}000$, and $p=4$.}
\label{fig:incore-twelve}
%\end{minipage}
\end{figure*}

We provide in Fig.~\ref{fig:incore-twelve} timings for both algorithms
when using 12 threads. As expected, the speedup of {\sc hp-gwas} with respect to 
{\sc gwfgls} soars from 8 to 50.

\section{Out-of-core algorithm}
\label{sec:ooc}

So far, we have developed an algorithm for GWAS % ({\sc hp-gwas})
that overcomes the limitations of current approaches. It
\begin{enumerate}
\item solves a sequence of GLS problems, 
\item exploits the available knowledge specific to GWAS, and
\item achieves high performance and scalability.
\end{enumerate}
However, the algorithm presents a critical limitation: data 
must fit in main memory. The most common scenarios of GWAS require the processing of
data sets that greatly exceed common main memory capacity:
in a typical scenario, where 36 millions of GLS problems are to 
be solved with $n=10{,}000$, the size of the input operand $\mathcal{X}_R$ %$m \times n$, 
is roughly 3 terabytes.
To overcome this limitation, we turn our attention to out-of-core 
algorithms~\cite{Toledo:1999:SOA:327766.327789}. The goal is to design
algorithms that make a proper use of available input/output (I/O) mechanisms
to deal with data sets as large as the hard-drive size, while
sustaining in-core high performance.

We regard the solution of GWAS as a process that takes as input a large stream of data, corresponding to
successive GLS problems, and generates as output
a large stream of data corresponding to the solution of such problems; 
thus, it demands out-of-core algorithms that efficiently stream data from secondary
storage to main memory and vice versa. 

We compare two approaches to data streaming. The first, used by GenABEL, is based on non-overlapping synchronous I/O;
because of wait states, this approach introduces a considerable overhead in the execution time.
The second, based on the well-known double-buffering technique, allows the overlapping of I/O with computation;
thanks to the overlapping, wait states, and the associated overhead, are 
reduced or even completely eliminated.

%\begin{figure*}
%\centering
\begin{minipage}[t]{\linewidth}
%\vspace{0pt}
\renewcommand{\algname}{}
\renewcommand{\lstlistingname}{Algorithm}
\begin{lstlisting}[caption=Out-of-core algorithm for  GWAS based on non-overlapping I/O, escapechar=!,label=alg:naive-ooc]
$L L^T = M$                 (!\sc potrf!)
$X_L := L^{-1} X_L$                 (!\sc trsm!)
$y := L^{-1} y$                 (!\sc trsv!)
$S_{TL} := X_L^T X_L$                 (!\sc syrk!)
$b_T := X_L^T y$                 (!\sc gemv!)
for each $X_{blk}$ in $\mathcal{X}_R$
  !\textcolor{red}{read}! $(X_{blk})$
  $X_{blk} := L^{-1} X_{blk}$                 (!\sc trsm!)
  for each $X_{R_i}$ in $X_{blk}$
    $S_{BL_i} := X_{R_i}^T X_L$                 (!\sc gemv!)
    $S_{BR_i} := X_{R_i}^T X_{R_i}$                 (!\sc dot!)
    $b_{B_i} := X_{R_i}^T y$                 (!\sc dot!)
    $b_i := S_i^{-1} b_i$                 (!\sc posv!)
  !\textcolor{red}{write}! $(b_{blk})$
\end{lstlisting}
%\makeatletter\def\@captype{lstlisting}\makeatother
\end{minipage}
%\hfill
%\begin{minipage}[t]{0.5\textwidth}
%\vspace{0pt}

\begin{figure}
\centering
\includegraphics[scale=.86]{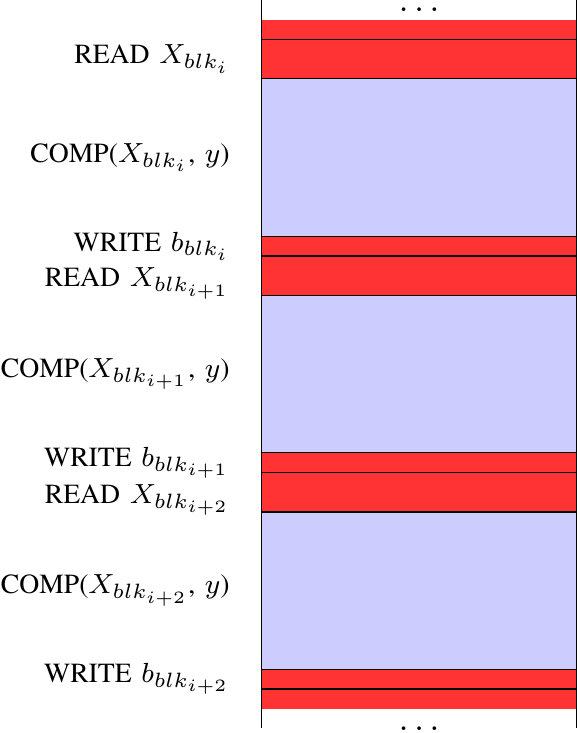}
%\makeatletter\def\@captype{figure}\makeatother
\caption{Non-overlapping approach to data streaming for out-of-core  GWAS.
I/O causes an overhead of 5\% to 10\%.} 
\label{fig:naive-ooc}
%\end{minipage}
\end{figure}

\subsection{Non-overlapping approach}
\label{subsec:naive}

The application of non-overlapping synchronous I/O to our in-core algorithm
({\sc hp-gwas}) results in Algorithm~\ref{alg:naive-ooc}. The algorithm
first computes the operations common to every GLS problem (lines 1 to 5) and then iterates
over the stream of $X_R$'s (lines 6 to 14). At each iteration, the following actions are performed:
\begin{enumerate}
\item read the $X_R$'s for a block of successive GLS problems,
\item compute the solutions, $b$'s, of such problems, and
\item write the $b$'s to disk.
\end{enumerate}
Both I/O requests (lines 7 and 14) are synchronous: after the requests are issued,
the processor enters a wait state until the I/O transfer has completed.
Fig.~\ref{fig:naive-ooc} depicts this shortcoming: red (dark) regions represent computation stalls where
the processor waits for data to be read or written; blue (light) regions represent actual computation.
Since loading data from secondary memory is orders of magnitude slower than loading data from
main memory, I/O operations introduce a considerable overhead that negatively impacts performance. 
For the scenario described above, in which $n=10{,}000$, $p=4$, and $m=36{,}000{,}000$, 
synchronous I/O applied to {\sc hp-gwas} causes a 5\% to 10\% overhead.

\subsection{Overlapping approach - Double buffering}
\label{subsec:doublebuff}

\begin{figure*}
\centering
%\begin{minipage}[t]{0.95\textwidth}
%\centering
\includegraphics[scale=.8]{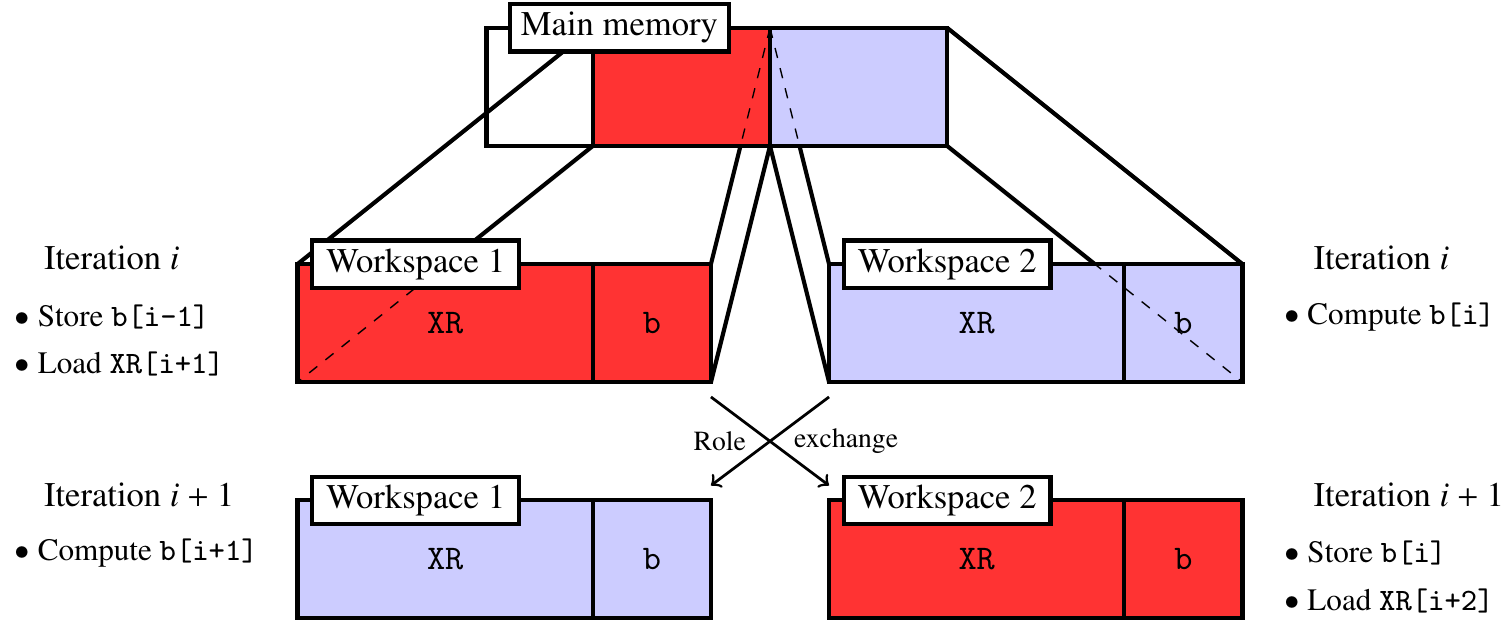}
\caption{Workspaces for double buffering. The main memory is divided, from left to right, in global data,
workspace 1, and workspace 2. Initially, workspace 1 is used for I/O and workspace 2 is used
for computation. After each iteration, the workspaces exchange roles.} 
\label{fig:workspaces}
%\end{minipage}
\end{figure*}

To put double buffering into practice, the main memory is split into two workspaces: one for downloading and uploading
data and one for computation. Also, the data streams are divided into blocks such that they fit in the corresponding
workspaces. While iterating over the blocks, the workspaces alternate their role, allowing the overlapping of I/O 
with computation, and reducing or even eliminating the overhead due to I/O.
Specifically for GWAS, both workspaces are subdivided in individual buffers, one for each 
operand to be streamed, $\mathcal{X}_R$ and $b$.
As illustrated in Fig.~\ref{fig:workspaces},
at iteration $i$, results from the previous iteration are located in {\tt Workspace1::b} and input
data for next iteration is to be loaded in {\tt Workspace1::XR}. Simultaneously, GLS problems corresponding to the current
iteration $i$ are computed and stored in {\tt Workspace2::b}.
At iteration $i+1$ the workspaces exchange their role: {\tt Workspace1} is used for computation, and 
{\tt Workspace2} is used for I/O.

It remains to be addressed how the downloading and uploading of data is actually
performed in the background while computation is being carried out. Our approach is based
on the use of asynchronous libraries, which allow a process to request
the prefetching of data needed for the next iteration: data is loaded in the background
while the process carries out computation with the current data set. 

In Algorithm~\ref{alg:double-buff} we provide the out-of-core algorithm {\sc ooc-hp-gwas} that
applies double-buffering to {\sc hp-gwas}.
At each iteration $i$ over the blocks of data (lines 6 to 16), the algorithm performs the following steps:
\begin{enumerate}
\item request the loading of the next block of input data ({\tt XR[i+1]}),
\item wait, if necessary, for the current block of data ({\tt XR[i]}),
\item compute the current set of problems defined by the current set of data,
\item request the storage of current results ({\tt b[i]}), and
\item wait, if necessary, until previous results are stored ({\tt b[i-1]}).
\end{enumerate}

\begin{figure}
\centering
\includegraphics{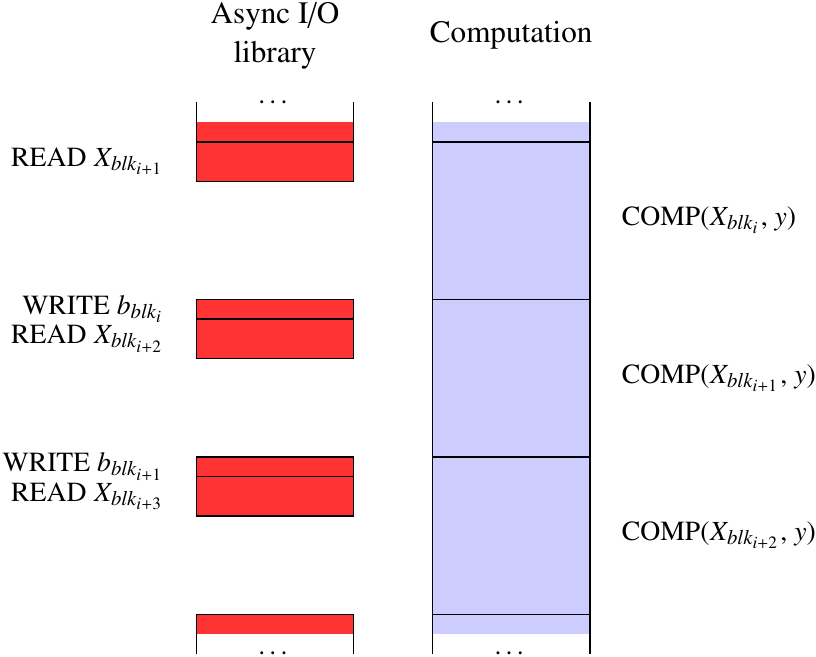}
%\makeatletter\def\@captype{figure}\makeatother
\caption{Overlapping approach to data streaming, based on an asynchronous library,
for out-of-core GWAS. The figure depicts a perfect overlapping of I/O with computation.} 
\label{fig:double-buff}
%\end{minipage}
\end{figure}

As illustrated in Fig.~\ref{fig:double-buff}, a perfect overlapping of I/O with computation 
means that no I/O is exposed and no processor idles waiting for I/O operations.

%\begin{figure*}
%\centering
\begin{minipage}{\linewidth}
%\vspace{0pt}
\renewcommand{\algname}{\textcolor{black}{$\ $({\sc ooc-hp-gwas})}}
\renewcommand{\lstlistingname}{Algorithm}
\begin{lstlisting}[caption=Out-of-core algorithm for GWAS based on double-buffering, escapechar=!,label=alg:double-buff]
$L L^T = M$                 (!\sc potrf!)
$X_L := L^{-1} X_L$                 (!\sc trsm!)
$y := L^{-1} y$                 (!\sc trsv!)
$S_{TL} := X_L^T X_L$                 (!\sc syrk!)
$b_T := X_L^T y$                 (!\sc gemv!)
for each $X_{blk}$ in $\mathcal{X}_R$
  async_read(next $X_{blk}$)
  !\textcolor{red}{wait}!(current $X_{blk})$
  $X_{blk} := L^{-1} X_{blk}$                 (!\sc trsm!)
  for each $X_{R_i}$ in $X_{blk}$
    $S_{BL_i} := X_{R_i}^T X_L$                 (!\sc gemv!)
    $S_{BR_i} := X_{R_i}^T X_{R_i}$                 (!\sc dot!)
    $b_{B_i} := X_{R_i}^T y$                 (!\sc dot!)
    $b_i := S_i^{-1} b_i$                 (!\sc posv!)
  async_write(current $b_{blk}$)
  !\textcolor{red}{wait}!(previous $b_{blk})$
\end{lstlisting}
%\makeatletter\def\@captype{lstlisting}\makeatother
\end{minipage}
%\hfill
%\begin{minipage}[t]{0.5\textwidth}
%\vspace{0pt}

\subsection{Sustaining in-core high performance}

A perfect overlapping is only one of two requirements for the out-of-core routine
to sustain in-core high performance. The second is to ensure that the operations within
the loop over the stream of data (lines 6 to 16) attain the same efficiency
as in the in-core routine. Both requirements depend on the number of threads and the block size, i.e., 
the number of $X_R$'s loaded at each iteration.

To completely eliminate the overhead due to data movement from disk to memory and vice versa,
the following equation must hold:
$$\nonumber {\tt time(computation) > time(load) + time(store)}.$$
The block size has to be large enough to ensure that for each iteration
the time spent in computing is larger than the time spent on storing the
previous results and loading data for the next iteration.
Since the computation time varies with the number of threads, the block
size needs to be adjusted accordingly.

Although it may seem that the best approach to select a block size
is to simply maximize memory usage, the initial overhead must be
taken into account: the loading of the first block of data is not
overlapped with computation. In systems equipped with large amounts
of main memory, it is advised to initiate the computation with a small
block size to avoid exposed I/O, and increase it after a few iterations.

\section{Performance results (II)}
\label{sec:results-ooc}

\begin{figure*}
\centering
\includegraphics[scale=0.83]{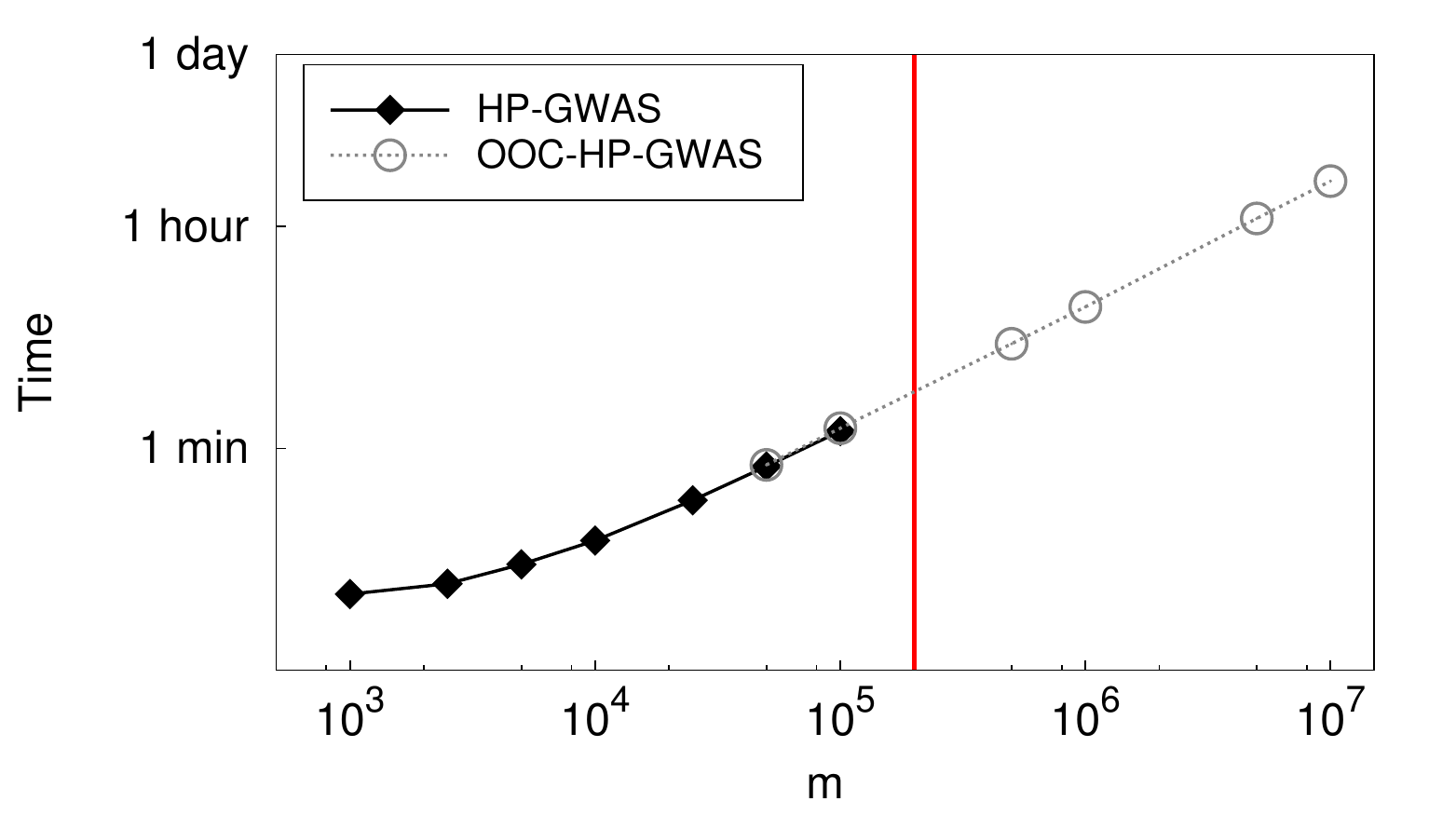}
\caption{Our out-of-core routine, {\sc ooc-hp-gwas}, sustains in-core performance for
problems as large as the available secondary storage of 1 terabyte.
The vertical line indicates the size limit for the in-core routine.
The results were obtained using 12 threads. The other problem dimensions 
are: $n=10{,}000$, and $p=4$. We used a block size of $5{,}000$ throughout.} 
\label{fig:ooc}
\end{figure*}

In this section, we focus on the experimental results for {\sc ooc-hp-gwas}, our out-of-core routine.
To measure the performance of the incorporated out-of-core mechanism,
we compare the timings with those of the in-core routine, previously
shown in Fig.~\ref{fig:incore-twelve}. We complete the picture with 
a comparison between the timings of our routines
and those of the in-core and out-of-core implementations of {\sc gwfgls}.
All routines were written in C, and the experiments
were run in the same environment as Section~\ref{sec:results-incore}.
In addition, {\sc ooc-hp-gwas} utilizes the AIO (asynchronous
input/output) library, available on UNIX systems as part of their standard libraries.

In Fig.~\ref{fig:ooc}, we combine timings for both the in-core routine {\sc hp-gwas} 
and the out-of-core routine {\sc ooc-hp-gwas}. 
The in-core routine is used for problems whose data sets fit in main memory,
and we switch to the out-of-core routine for larger problems. 
The vertical line indicates the size of the largest problem that can be solved in-core.
The figure shows that, thanks to the double-buffering and an appropriate
choice of the block size, {\sc ooc-hp-gwas} achieves a perfect
overlapping of I/O with computation. As a consequence,
{\sc ooc-hp-gwas} is able to sustain in-core performance for problems as large as the hard-drive size. %, in this case 1 terabyte. 

In Table~\ref{tab:ooc-ratios}, we collect timings for both our routines and {\sc gwfgls} 
in both in-core and out-of-core scenarios. 
The provided ratios confirm the impact of using a sub-optimal approach to out-of-core:
While, as seen in Fig.~\ref{fig:ooc}, the overlapping I/O mechanism incorporated in {\sc ooc-hp-gwas} sustains in-core
performance, the non-overlapping approach in {\sc gwfgls} results in a 10\% to 15\% overhead.
As a consequence, the speedup over {\sc gwfgls} raises from 50 to 58.

\begin{table*}
\renewcommand{\arraystretch}{1.4}
\centering
\footnotesize
\begin{tabular}{ l | c c c || c c c c } \toprule
	  \multicolumn{1}{r|}{$m = $} & $10{,}000$ & $50{,}000$ & $100{,}000$ & $500{,}000$ & $1{,}000{,}000$ & $5{,}000{,}000$ & $10{,}000{,}000$ \\\midrule
	  {\sc gwfgls}   & 429.2 & $2{,}072.5$ & $4{,}117.9$ & $24{,}065$ & $48{,}130$ & $240{,}650$ & $481{,}300$ \\
	  {\sc *hp-gwas} & 10.9 & 43.0 & 82.6 & 414 & 816 & $4{,}184$ & $8{,}343$ \\
	  Ratio $\frac{\text{\sc gwfgls}}{\text{\sc *hp-gwas}}$ & 39.2 & 48.1 & 49.9 & 58.1 & 58.9 & 57.5 & 57.7 \\
	\bottomrule
  \end{tabular}
\caption{Timings for  {\sc gwfgls} and {\sc *hp-gwas} ({\sc hp-gwas} and {\sc ooc-hp-gwas}) 
	for both in-core and out-of-core scenarios. 
	The problem dimensions are: $n = 10{,}000$ and $p = 4$, for an increasing value of $m$.
	The results were obtained using 12 threads. The timings are in seconds.
	The double vertical line separates timings for the in-core (left) and out-of-core (right) routines.
	The increase in speedup in the out-of-core case reflects the overhead introduced in {\sc gwfgls} due to a non-overlapping I/O.}
\label{tab:ooc-ratios}
\end{table*}

The largest tests presented, involving 10 millions of genetic markers ($X_R$'s), 
took less than 2.5 hours with {\sc ooc-hp-gwas}. This means that a complete genome-wide scan of association
between 36 millions of genetic markers in a population of $10{,}000$ individuals
takes now slightly more than 8 hours, and is only limited by the availability
of a large (and cheap) secondary storage device.

\section{Future Work}
\label{sec:future}

As current biomedical research experiences a large boost in the amount of available genomic data,
computational biologists are eager to solve problems of ever increasing size.
Even though the time spent to perform genome-wide analysis is reduced significantly thanks to
the techniques presented in this paper, further speedups are required to satisfy
future needs.

Throughout the paper we assumed, as is the case in current analyses,
that the matrix $M$ fits in main memory. In this context, a further reduction
of execution time can be achieved through distributed-memory architectures
via an MPI parallelization:
the processes would first perform the Cholesky factorization of $M$ redundantly, and
then operate on distinct chunks of $X_R$. In addition, if GPUs with enough
memory to host $M$ were available, the routines could be further sped up
by offloading the computation of the {\sc trsm} (line 8 in Alg.~\ref{alg:double-buff})
onto the devices.

When instead $M$ does not fit in main memory, one should rely on approaches based on
out-of-core algorithms-by-blocks~\cite{Quintana-Orti:2009:PMA:1527286.1527288,QuintanaOrti:2012:RTS}
and libraries such as ScaLAPACK~\cite{slug} and Elemental~\cite{elemental-toms},
respectively for shared-memory and distributed-memory architectures.

\section{Conclusions}
\label{sec:conclusions}

We tackled a problem, extremely common in bioinformatics, that
requires both high-performance computing and storage. Neither general
nor domain-specific libraries provide a viable solution. Indeed, due to the expected execution time
and storage requirements, it was believed that these problems could be
solved exclusively with the aid of supercomputers. This paper instead
demonstrates that a single multi-core node suffices.

We presented high-performance algorithms, and their corresponding implementations,
for the solution of sequences of generalized least-squares problems (GLSs) in the context
of genome-wide association studies (GWAS). When compared to the widely used {\sc gwfgls}
routine from the GenABEL package, our routines attain speedups larger than 50.

Our routines are specifically tailored for multi-threaded architectures.
We followed an incremental approach: starting from an algorithm to solve one single
GLS, we detailed the steps towards a high-performance algorithm for GWAS.
At each step, we identified the limitations of current existing libraries and tools,
and described the key insight to overcome such limitations.

First, we showed that no matter how optimized is a routine to solve a single
GLS instance, it cannot possibly compete with tools specifically designed
for GWAS: it is imperative to take advantage of the sequence of correlated
problems. This discards the black-box approach of traditional libraries.

Then, we identified {\sc gwfgls}' issues regarding efficiency, scalability,
and data handling, and detailed how we addressed them. 
Taking advantage of
problem symmetries and application-specific knowledge, we were able to
completely eliminate redundant computation. Next, we pointed out that
even BLAS-3 kernels might suffer from low efficiency. A careful rearrangement of
the operations leads to an efficiency of 94\%. Combining two kinds of parallelism
--a multi-threaded version of BLAS and OpenMP parallelism--, our in-core solver
attains speedups close to 11 on 12 cores.

Finally, thanks to an adequate utilization of the double-buffering technique, allowing for
a perfect overlapping of data transfers with computation, our out-of-core routine 
not only inherits in-core efficiency and scalability, but it is also 
capable of sustaining the achieved 
high performance for problems as large as the available secondary storage.

As an immediate result, our routines enable genome-wide association studies of 
unprecedented size and shift the limitation from computation time to size of 
secondary storage devices.

\section*{Acknowledgments}

Financial support from the 
Deutsche Forschungsgemeinschaft (German Research Association) through
grant GSC 111 is gratefully acknowledged. 
Y. Aulchenko was funded by grants from the Russian Foundation of Basic Research (RFBR), the Helmholtz society (RFBR-Helmholtz Joint Research Groups), and the MIMOmics project supported by FP7.
The authors thank Matthias Petschow for discussion on the algorithms, and
the Center for Computing and Communication at RWTH Aachen for the computing resources.

% ----------------------------------------------------------------

\bibliography{bibliography}
\bibliographystyle{acmtrans}

\begin{received}
Received Month Year;
revised Month Year;
accepted Month Year
\end{received}
\end{document}